\documentclass[%
 preprint,
showpacs,preprintnumbers,
 amsmath,amssymb,
 aps,
prb,
]{revtex4-1}

\usepackage{graphicx}
\usepackage{dcolumn}
\usepackage{bm}

\begin{document}

\title{Isotropic properties of the photonic band gap in quasicrystals with
low-index contrast}

\author{Priya Rose T.}
 \affiliation{CNR-SPIN and Department of Physics, University of Naples Federico
II, Napoli, Italy}
\author{E. Di Gennaro}
 \affiliation{CNR-SPIN and Department of Physics, University of Naples Federico
II, Napoli, Italy}
\author{G. Abbate}
 \affiliation{CNR-SPIN and Department of Physics, University of Naples Federico
II, Napoli, Italy}
\author{A. Andreone}
 \affiliation{CNR-SPIN and Department of Physics, University of Naples Federico
II, Napoli, Italy}
\email{andreone@unina.it}

\date{\today}

\begin{abstract}
We report on the formation and development of the photonic band gap in
two-dimensional 8-, 10- and 12-fold symmetry quasicrystalline lattices of 
low index contrast. Finite size structures made of dielectric
cylindrical rods were studied and measured in the microwave region, and their
properties compared with a conventional hexagonal crystal. Band gap
characteristics were investigated by changing the direction of propagation of
the incident beam inside the crystal. Various angles of incidence from $0^\circ$
to $30^\circ$
were used in order to investigate the isotropic nature of the band gap. The
arbitrarily high rotational symmetry of aperiodically ordered structures could
be practically exploited to manufacture isotropic band gap materials, which are
perfectly suitable for hosting waveguides or cavities.
\end{abstract}

\maketitle


\section{Introduction}
Structures exhibiting photonic band gap (PBG) characteristics are useful in
confining and guiding electromagnetic energy.  Photonic crystals (PCs) are
artificially engineered materials with spatially modulated refractive indices
that are widely used for such purposes. There is a tremendous interest in
studying their properties because of their potential application in design and
manufacturing of new optical components and devices like wavelength division
multiplexers, switches, light emitting diodes and lasers\cite{Potter, Hitoshi,
Fan, Park}.

Recently, structures lacking long-range translational order but with
orientational order and higher order rotational symmetries which are not
compatible with the spatial periodicity, called photonic quasi-crystals (PQCs),
are also gaining attention because of their unique characteristics. Photonic
quasicrystals have neither true periodicity nor translational symmetry, however
they can exhibit symmetries that are not achievable by conventional periodic
structures. These features have recently attracted a large interest because of
their potential impact in engineering novel optical circuits. PQCs can show
directive emission\cite{Micco}, mode confinement\cite{DiGennaro},
superlensing\cite{DiGennaro2}, as photonic crystals do. 

One-dimensional photonic band-gap has been observed in dielectric multilayers
stacked according to Fibonacci series\cite{Sibilia}. Dielectrics arranged
according to quasi-periodic geometries such as octagonal (8-fold), decagonal
(10-fold) and
dodecagonal (12-fold) are shown to have two-dimensional PBG\cite{Kaliteevski,
Kaliteevski2, Zhang, Jin}. Penrose-tiled (10-fold rotational symmetry) PQCs were
the most studied structures among those
presented above, and there were numerous studies about the mechanism of
formation of PBG\cite{DellaVilla}, optical properties,
diffraction pattern\cite{Kaliteevski} and multiple scattering\cite{Zhang} for
this geometry. An organic laser based on Penrose-tiled PQCs was also
demonstrated\cite{Notomi}. Formation of complete PBG in 12-fold symmetric
dodecagonal PQC was numerically and experimentally studied by Zoorob et
al.\cite{Zoorob}. 

The lack of periodicity renders the study of PQCs very complex and computationally demanding. Although some concepts developed for periodic PCs can be used for the analysis of the properties of PQCs, a rigorous extension of a Bloch-type theorem (and associated tools and concepts) does not exist. Previous methods include a “supercell” approach in a plane-wave expansion technique to approximate the response of the infinite aperiodic lattice \cite{Kaliteevski, Kaliteevski2}, or the use of Archimedean-like tilings with properties similar to those of photonic quasi-crystals \cite{Lourtioz}. Recently, a method was proposed to computing the spectra and the eigenstates of a PQC by directly solving Maxwell equations in a periodic unit cell of a higher-dimensional lattice \cite{Rodriguez}.

In the geometries mentioned above, the tiling is previously calculated by
matching rules or inflation/deflation algorithms or using projection methods
from a hypercubic lattice\cite{Steurer}. Well-known tilings are the octagonal
Ammann-Beenker\cite{Socolar}, the decagonal Penrose\cite{Penrose}, and the
dodecagonal pattern based on the Stampfli rule\cite{Stampfli}, which represent
the quasiperiodic structures object of the present study.  

The interference patterns formed by multiple-beam interferometry provide another
way to obtain highly symmetric quasicrystalline patterns\cite{Wang}. Recently,
holographic techniques based on interferential methods are used to fabricate
PQCs with very high rotational symmetry, the fold being determined by the number
of interfering beams. The PBG properties of 12-fold symmetric
quasi-crystal patterns formed by double-beam multiple exposure holography were
studied theoretically by Gauthier et al.\cite{Gauthier}. 

The method of single-beam computer-generated holography has been also successfully used to
fabricate PQC structures with up to 23-fold rotational symmetry\cite{Zito}. These techniques can be used to produce complicated two-dimensional geometries with ease. 
Very recently, a three-dimensional Penrose-type PQC fabricated through this method was also reported\cite{Harb}. In considering holographic lithography, it is important to have geometries possessing optimum band-gap properties at low refractive index contrast. A conceivable application is the combination of this versatile technology and soft materials like Polymer Dispersed Liquid Crystals for the realization of large area, high quality, low cost optical devices with switchable properties\cite{Gorkhali}.

In two-dimensional structures, a full PBG for both transverse
magnetic (TM, $\mathbf{E}$ field orthogonal to the crystal plane) and transverse electric
(TE, $\mathbf{H}$ field orthogonal to the crystal plane) polarization is possible for high
values of the index contrast only. Nevertheless, the minimum refractive index
contrast $\Delta n$\cite{ind_con} at which a partial band gap starts to appear, for a specifically polarized electromagnetic mode, is an important parameter for the development of photonic devices such as waveguides. This is very much dependent on the particular geometry used. 
\begin{figure*}
\includegraphics[width=0.75\textwidth]{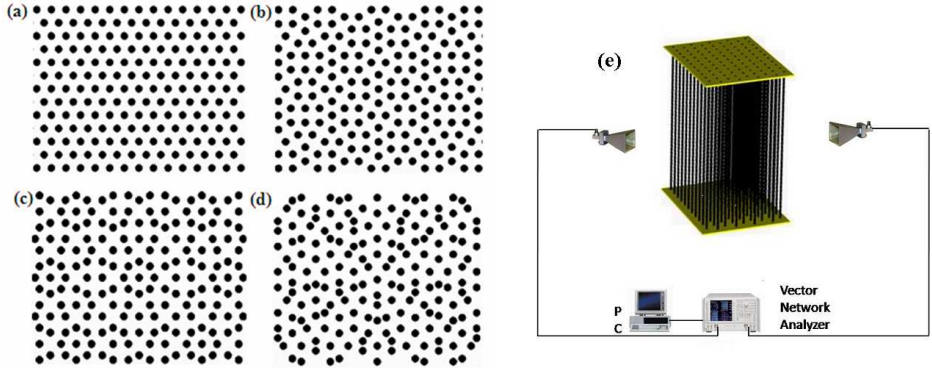} 
\caption{The periodic and aperiodic geometries under study: (a) hexagonal, (b) dodecagonal, (c) octagonal, and (d) decagonal (Penrose). (e): schematic diagram of the experimental set-up (not in scale). Horn antennas are used to transmit and receive the microwave radiation.  Data are collected using a  vectorial network analyzer computer-controlled. The size of each structure
(periodic or aperiodic) is $40\times 14\times 60 cm^3$}\label{fig1}
\end{figure*} 

The minimum value for opening a partial band gap in a periodic (triangular)
structure was calculated using a plane wave technique by Matthews et
al.\cite{Matthews}. For the optimal configuration of dielectric circular rods in air (with a filling factor - the ratio between total and rods occupied area - $\eta=0.3$), the critical value of the dielectric permittivity is $1.73$ for the TM band gap (corresponding to $\Delta n =0.31$), much larger for the TE band gap. This study should rule out any attempt to develop PBG devices based on periodic structures with a lower index contrast.

In the case of quasi-crystal geometries, a novel method based on density waves
was used to determine the polarized band gap for $n-fold$ rotational symmetrical
structures as a function of the index contrast\cite{Rechtsman}. This method
works well for TM  polarization only, for which the density function for the optimal  configurations tends to have smoother features (on the scale of the wavelength) compared to the case of TE polarization. At low contrast, quasicrystalline structures of high
symmetry tend to have larger band gaps than the crystalline ones, but smaller
band gaps at high contrast. The sixfold (hexagonal) crystalline structure yields
the largest gap for any value of the index contrast approximatively higher than
$1$. The critical value for the optimized quasi-periodic photonic crystals is estimated to be close to $\Delta n =0.22$.  

Very recently, a detailed numerical study on the band-gap formation at low
index-contrasts for both TE and TM polarization was carried out by Zito et
al.\cite{Zito2}. Using Finite Difference Time Domain (FDTD) simulations, authors
showed that, more than the degree of rotational symmetry, the difference in the tiling geometry might dramatically affect the existence and behavior of the band-gap. 

Another critical parameter that comes into play for the realization of devices
is the isotropy of the PBG. For some applications, such as light
emitting diodes, this property may be desirable, even if the size of the full
gap is slightly reduced.  The periodic structures with square or triangular lattices have anisotropic band-gap properties because of the anisotropy of the Brillouin Zone (BZ). Significantly, optimized quasicrystal gaps are more isotropic than those of
crystals, for all contrasts, due to their disallowed rotational symmetries. This
is due to the fact that their effective BZs are more circular than the BZs of
the periodic structures, which translates in reduced frequency undulations (that
is, band gap variations over the wavevector in the BZ). 

Experimental investigations in this direction were performed by Bayindir et al.\cite{Bayidir} in the microwave regime and by Hase et al.\cite{Hase} in the far infrared region, based on octagonal and Penrose quasicrystals. Both studies reported the appearance of a PBG having almost isotropic properties in aperiodic lattices consisting of dielectric rods in air and for electric field parallel to the rods.
  
We present here a detailed and systematic analysis of the band-gap isotropy of
photonic crystal and quasi-crystal structures having low refractive index
contrast and for both TM and TE field polarization. We consider four different
geometries for comparison, the periodic hexagonal pattern with $6-fold$ symmetry
(Fig. \ref{fig1}(a)), and quasi-crystalline geometries with $12-fold$ symmetry
(dodecagonal), $8-fold$ symmetry (octagonal) and $10-fold$ symmetry (Penrose decagonal) 
as shown in Fig. \ref{fig1}(b), Fig. \ref{fig1}(c) and Fig. \ref{fig1}(d) respectively. 
The PBG properties are studied both numerically and experimentally for two different 
index contrasts close to the critical values.

\section{Computational and Experimental Methods}
The photonic crystals studied consist of infinitely long dielectric cylindrical rods in air placed on the vertices of tiles in the corresponding geometry. The filling
factor $\eta$ is set to be the same for all the structures under study, and is equal to
$0.23$. The geometries used are shown in Fig. \ref{fig1}. They are designed to
have approximately $400$ rods in an area of $\sim 40 a$ x $ 14 a$, where $a$ is a characteristic length of the same order of the tile side length of each structure.

As discussed in the introduction, from the computational point of view each quasi-crystalline structure presents a challenge in obtaining the information on the photonic band-gap since the lack of translational symmetry prevents the rigorous use of the Bloch theorem in the calculations. This difficulty has hampered the use of supercell or similar techniques in the modeling of real aperiodic structure.

The Finite Difference Time Domain (FDTD) technique is useful in this respect, since it can provide an alternative, fast and accurate method to simulate in the real space the propagation of the electromagnetic waves through a finite portion of a quasi-crystal without recurring to any “approximant”.  2D FDTD method with uniaxial perfectly matched layer (PML) boundary conditions (along x- and y-directions) is employed to obtain transmission characteristics as a function of frequency of the incident radiation, propagation direction and polarization. In each simulation, the source of excitation consists of a time-pulsed Gaussian beam (with $250a < b <800a $, where $b$ is the confocal parameter) placed outside the crystal structure and impinging on it with different incidence angles. The field components after propagation through the crystal are collected using a detector (time monitor) placed on the other side of the crystal. The Fourier transform of these data gives the transmission properties as a function of frequency. Both TM and TE polarizations were analyzed in the simulations.

The experiment was designed and carried out in the microwave regime ($9-20 GHz$).
Cross-linked polystyrene (Rexolite) and polytetrafluoroethylene (Teflon)  having radius $0.64 cm$ and length $60 cm$ were used to fabricate the cylindrical rods. These materials show a dielectric constant of $2.56$ and $2.1$ respectively in the frequency region of interest and a relatively low dissipation. Loss tangent values are in the range between $10^{-3}$ and $10^{-4}$. In order to build up the structure, circular holes with the designed geometries have been drilled onto two  support dielectric plates, fixed $60 cm$ apart and then filled with the rods. The characteristic length $a$ is chosen to be of the order of $1 cm$, so that the first PBG appears in the region of $10-12 GHz$ for all structures under test.

Two high-gain horn antennas acting as transmitter and receiver and connected to a two-port
vectorial network analyzer (VNA) $HP 8720C$ have been used to obtain the transmission characteristics of the crystals. To change the field polarization, both horn antennas are rotated by $90^\circ$. The distance between transmitter and receiver is set to $3m$, so that the polarized signal can be considered a plane wave. In the frequency region of interest, the signal wavelength is much smaller than the crystal height in the z-direction, therefore one can safely assume that the response has a two dimensional character in the crystal $(x-y)$ plane. Before each measurement the VNA is calibrated in absence of the material under study. Transmission curves are then obtained by introducing the sample in between the two horn antennas. 

In order to study the isotropic nature of the PBG, the transmission spectra are obtained as a function of incidence angle for all geometries under study. To record the transmission as a function of direction of propagation of the incident radiation, the crystal is rotated in $5^\circ$ steps in respect to the normal direction while keeping the position of the antennas unchanged. Because of the $n-fold$ rotational symmetry of the structures under test, the crystal/quasi-crystal properties need only be examined over the $5^\circ- 180^\circ/n$ range since they repeat themselves for any propagation angle outside these degree values. For the sake of clarity, not all angles will be shown in the spectra of the different structures. Measurements were also carried out as a function of crystal thickness, for normal transmission only.

\begin{figure*}
\includegraphics[width=0.9\textwidth]{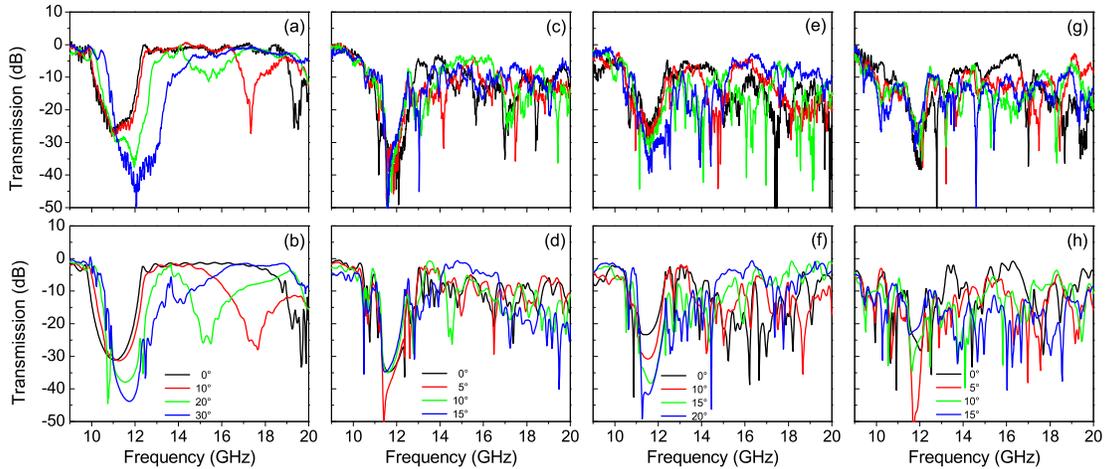} 
\caption{Experimental and calculated transmission spectra of the
hexagonal [(a) and (b)], dodecagonal [(c) and (d)], octagonal [(e) and (f)], and decagonal [(g) and (h)] photonic crystals respectively, for TM polarization. Curves of different colours correspond to different angles of incidence as indicated in the graph.}\label{fig2}
\end{figure*} 

\section{Results and Discussion}
The transmission characteristics are measured as a function of incidence angle and compared with simulation results for all geometries under study, using the two materials with different index contrast ($\Delta n = 0.6$ and $0.44$ for Rexolite and Teflon respectively) and for both TM and TE polarization.
\subsection{Refractive index contrast 0.60}
\noindent
\textbf{TM polarization:}\\
Fig. \ref{fig2} shows the experimental and calculated transmission spectra
for all the structures under study with an index contrast of $0.60$ for various
angles of propagation. The experimental   results for the hexagonal photonic crystal are
presented in Fig. \ref{fig2}(a) whereas Fig. \ref{fig2}(b) shows the
corresponding simulation data. In all curves there is a strong attenuation, from around $30$ to $40 dB$ and more, in the transmission of the electromagnetic waves through the crystal in the region where the first PBG is present. From the graphs it is clearly observed that the
transmission spectra change drastically as the angle of propagation is changed from $0^\circ$ to $30^\circ$. In the case of normal signal incidence ($0^\circ$), the region of
low transmission (photonic band-gap) is centered at about $11 GHz$ and spans for $\sim 2.6 GHz$, with almost no change for an angle $10^\circ$. However, as the angle is increased to
$20^\circ$, changes are clearly visible. The center of the PBG is shifted to $12 GHz$ whereas its width becomes larger ($\sim 3.3 GHz$). For an angle of $30^\circ$, the widening of the PBG is even stronger. Attenuation in the PBG frequency region increases as the incidence angle does, because of the stronger diffraction light undergoes due to the longer path inside the finite crystal. For the first bandgap, experimental results matches very well the numerical data reported in Fig. \ref{fig2}(b). At higher frequencies, a second bandgap appears with less pronounced characteristics but with similar angular dependence. Also in this case there is a fairly good agreement between measurements and simulations. 
  
The transmission characteristics of a dodecagonal PQC structure are shown in Figs. \ref{fig2}(c) and (d). A well-pronounced PBG, characterized by a signal attenuation larger than $40 dB$ and extending approximately from $11$ to $12.6 GHz$, is observed above the valence band (lower band) frequencies. At higher frequencies (upper bands) the transmitted power decreases, possibly with the presence of smaller and shallower bandgaps up to the maximum measurement frequency of $20 GHz$. The experimental (Fig. \ref{fig2}(c)) and simulation (Fig. \ref{fig2}(d)) angular dependence puts in evidence that in this case the PBG between the valence and conduction bands is nearly isotropic. Transmission spectra are only slightly affected by the change of the angle of propagation from $0^\circ$ to $15^\circ$. 

Similar features are observed for the other aperiodic photonic crystals. In the case of the $8-fold$ and $10-fold$ PQCs, the distribution of cylinders has a mirror symmetry with respect to the line of $22.5^\circ$ and $18^\circ$ respectively, and the data are reported in the angular range $0-20^\circ$

The numerical and experimental results for the case of the octagonal geometry
are presented in Figs. \ref{fig2}(e) and \ref{fig2}(f) respectively.
The PBG is centered at $11.5 GHz$ for angles from $0^\circ$ to $10^\circ$ whereas it is shifted to $\sim11.8 GHz$ for incidence at larger angles. There is also a small variation in the width of the PBG, from $1.4 GHz$ to $\sim 2 GHz$.  

Data obtained for the Penrose tiled quasicrystal are shown in Figs. \ref{fig2}(g) and (h). From both measurements (Fig. \ref{fig2}(g)) and simulations (Fig. \ref{fig2}(h)) one can clearly see that the bandgap is not as deep and wide as in the other cases, nevertheless its position and width remains almost the same for all angles. 

As in the periodic case, the transmitted signal shows an increasing attenuation in the PBG as a function of the incidence angle for all the aperiodic structures.  Not surprisingly, the geometry with $12-fold$ symmetry is the less sensitive to the change in the signal propagation direction. This is likely related to the fact that the bandgap for the dodecagonal pattern may be associated with a more short range ordering of the dielectric scattering centers in comparison with the other aperiodic structures \cite{Chan}.
  
\begin{figure}
\includegraphics[width=0.45\textwidth]{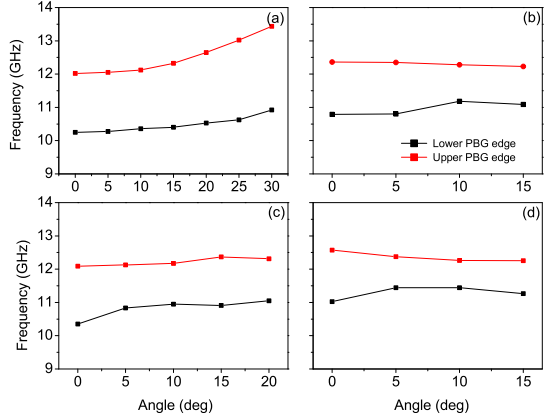} 
\caption{Variation of PBG for TM polarization as a function of angle for the different geometries: (a) hexagonal, (b) dodecagonal, (c) octagonal and (d) decagonal. In each graph, the black and the red curves indicate the lower and upper frequency edges of the band gap respectively.}\label{fig3}
\end{figure}

In order to better determine the directional variation of PBG, we measured the upper and lower PBG boundaries as a function of the propagation angle. The band gap edges are defined as the frequency values for which attenuation reaches $15 dB$. Data in Fig. \ref{fig3} summarize the results found from the observation of the angular dependence. The hexagonal geometry (Fig. \ref{fig3}(a)) clearly shows a strong dependence of the upper and lower frequency edges as a function of angle, whereas the PBG width seems to be less affected. The results for the dodecagonal structure, instead, clearly indicates (see Fig. \ref{fig3}(b)) that its response is highly isotropic, with very small variations of the PBG width and edges at different angles from $0^\circ$ to $15^\circ$. The properties for the octagonal geometry lie somehow in between, since it shows less isotropy compared to the dodecagonal geometry and a noticeable width dependence in respect with the hexagonal case, as displayed in Fig. \ref{fig3}(c) between $0^\circ$ and $20^\circ$. Penrose geometry also seems to have quite isotropic, but narrower PBG (Fig. \ref{fig3}(d)) in the same incidence angle range.
If the transmission at normal incidence is considered, the hexagonal crystal shows the widest PBG compared to all other structures under study. The PBG for the periodic crystal is around $2.6 GHz$. In the case of aperiodic structures, this value decreases from $2.2 GHz$ (octagonal) to $2 GHz$ (dodecagonal) down to $1.1 GHz$ (Penrose).
\\

\begin{figure}
\includegraphics[width=0.45\textwidth]{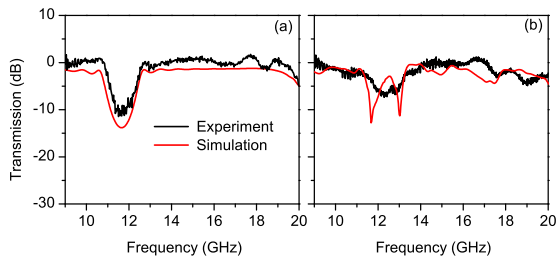} 
\caption{The transmission curves for (a) hexagonal and (b) dodecagonal
geometries for normal incidence and TE polarization. Both experimental and simulated results are
shown in each graph (red and black curves respectively).}\label{fig4}
\end{figure}

\noindent
\textbf{TE polarization:}\\
The results obtained for the hexagonal and dodecagonal geometries are compared in Fig. \ref{fig4} for normal incidence only. The hexagonal geometry (Fig. \ref{fig4}(a)) clearly shows a PBG $\sim 1.7 GHz$ wide centered at about $11.7 GHz$. The experimental (black curve) and calculated (red curve) results are in good agreement. The dodecagonal structure (Fig. \ref{fig4}(b)) displays a very shallow bandgap measured approximately in the region $12-13 GHz$ (black curve), however only partially reproduced by numerical data (red curve).

Simulation of the response to a TE polarized wave shows instead that the transmission characteristics for the Penrose and octagonal structures are nearly featureless\cite{Zito2}  and are not reported here.

\subsection {Refractive index contrast 0.44}

The results presented in this case are for the hexagonal and dodecagonal cases only. The case of $8-fold$ and $10-fold$ were numerically examined in a previous paper\cite{Zito2}, where we found that these geometries does not show any clear bandgap behavior in the frequency region of interest for index contrast values below $0.6$ and for both polarizations. 
\begin{figure}
\includegraphics[width=0.45\textwidth]{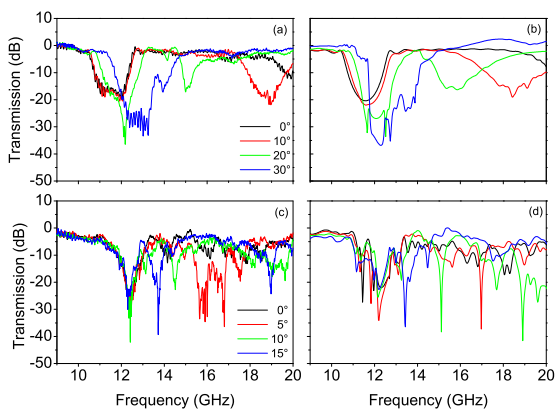} 
\caption{ Experimental and  calculated transmission spectra of the
hexagonal [(a) and (b)], and dodecagonal [(c) and (d)] photonic crystals respectively, with an index contrast of 0.44 for TM polarization. Curves of different
colours correspond to different angles of incidence as indicated in the
graph.}\label{fig5}
\end{figure} 

\begin{figure}
\includegraphics[width=0.45\textwidth]{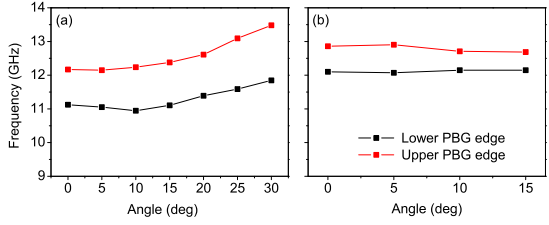} 
\caption{Variation of PBG as a function of angle for (a) hexagonal and (b)
dodecagonal geometries, for TM polarization. In each graph, the black and the red curves indicate the
lower and upper frequency edges of the band gap respectively.}\label{fig6}
\end{figure}

For TM polarization, the observed features are similar to the case of the higher refractive index contrast presented above. In this case too, the hexagonal crystal shows a clear variation in the PBG properties as the angle of propagation is varied, as seen in Figs. \ref{fig5}(a) (experiment) and \ref{fig5}(b) (simulation). Similar results for the dodecagonal crystal are presented in Figs. \ref{fig5}(c) and \ref{fig5}(d) respectively. The PBG frequency shift is reduced in comparison with the hexagonal crystal but the PBG is much narrower. A graph showing the variation of the upper and lower band-edge frequencies as a function of angle is also plotted in this case and shown in Fig. \ref{fig6} for both the $6-fold$ and $12-fold$ symmetry (in the range of $0-30^\circ$ and $0-15^\circ$ respectively). 
As expected due to the lower index contrast, using Teflon cylinders the observed PBGs are never as pronounced as in the case of Rexolite.

For TE polarization, the PBG features shown in the spectra were so weak to render the study of its angular dependence meaningless.  Once again, this is in agreement with the previous simulation work\cite{Zito2}.

\begin{figure}
\includegraphics[width=0.45\textwidth]{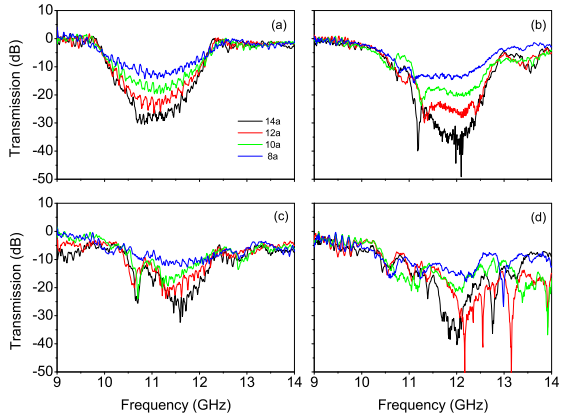} 
\caption{Experimental transmission curves as a function of sample thickness for (a) hexagonal, (b) dodecagonal, (c) octagonal, and (d) decagonal geometries with  index contrast $0.60$  and TM polarization.}\label{fig7}
\end{figure} 

\subsection{Effects of thickness}
To understand how sensitive is the PBG to the size of a photonic quasi-crystal with a low index contrast, transmission spectra were also measured by removing the cylinder rows                                                                                                                                                                                                                                                                                           simultaneously from both sides of the structure in the signal propagation direction. Results are shown in Fig. \ref{fig7} for all aperiodic geometries with an index contrasts of 0.6 for normal incidence and in the bandgap region only. The hexagonal case is also reported for comparison. 
Graphs clearly show that the dodecagonal pattern is the less affected by the reduction in size, presenting an appreciable bandgap ($15 dB$ attenuation) even when the slab thickness decreases from $14a$ to $8a$. Similar results, but obviously less pronounced, are also obtained for Teflon ($10dB$ attenuation). 
This is a further confirmation that the 12-fold geometry is the most robust amongst the different structures under investigation because of the stronger role played by the short-range interactions, and it is therefore preferable for the development of very compact photonic devices.

\section{Conclusions}
It is well known \cite{Joann} that for transverse magnetic modes bandgaps can be easily achieved in photonic crystals made of isolated dielectric rods, whereas connected lattices favour transverse electric gaps. The same happens in photonic quasi-crystals, where we have experimentally shown that in the dielectric-in-air configuration two-dimensional TM bandgaps are clearly visible whereas the response of the aperiodic structures to a TE polarised signal is very weak or almost featureless.
Besides that, results put in evidence that TM bandgaps in PQC are possible even with a very low-index-contrast, in agreement with numerical studies carried out in Ref. \onlinecite{Rechtsman}. In particular, well-pronounced bandgaps are present for the dodecagonal geometry, as theoretically expected because of the higher symmetry. 
 
Also, we did observe that gaps in quasicrystalline geometries are more isotropic - although narrower - than those in periodic crystals, due to their disallowed, non-crystallographic, rotational symmetries. That is, the position and width of the PBG are almost independent of the incident angle of the light, contrarily to their periodic counterparts, where gaps of different directions may appear at different frequencies because of non-spherical first BZ. For specific applications, like light emitting diodes or waveguides, very isotropic PBGs may be desirable, even if the size of the full gap is slightly reduced. 

Another important feature of PQCs is that the existence of the gaps is governed by the short-range environment. This is particularly evident in the case of the dodecagonal geometry, where the PBG is robust even for a significant reduction in size in the propagation direction. 

In conclusion, this study confirms that quasicrystalline structures having long-range orientational order forbidden for periodic systems are promising candidates as PBG materials. Photonic crystals based on aperiodic specific geometries present extremely interesting features that cannot be achieved in the periodic case. The low index contrast allows the use of versatile and low-cost technologies like holographic lithography combined with soft materials for the development of compact devices with switchable properties for an all-optical ultrasmall integrated circuitry.


\begin{thebibliography}{34}
\bibitem [{ind()}]{ind_con}%
  \BibitemOpen
  \href@noop {} {}\bibinfo {note} {The index contrast is defined in the
  following simply as the difference between the refractive indexes of the
  dielectric material and air}\BibitemShut {NoStop}%
\end{thebibliography}
\end{document}